\documentstyle[12pt,epsfig]{article}
\def\beq{\begin{equation}}
\def\eeq{\end{equation}}
\def\be{\begin{eqnarray}}
\def\ee{\end{eqnarray}}
\def\bea{\begin{eqnarray}}
\def\eea{\end{eqnarray}}

\newcommand{\gsim}{\lower.7ex\hbox{$\;\stackrel{\textstyle>}{\sim}\;$}}
\newcommand{\lsim}{\lower.7ex\hbox{$\;\stackrel{\textstyle<}{\sim}\;$}}

\begin{document}

\begin{center}
\vspace{-3ex}{
                      \hfill hep-ph/0405088}\\[2mm]

\vspace{1.0cm}

{\Large \bf
Neutrino Mixing and Quark-Lepton Complementarity}

\vspace{0.6cm}

H.~Minakata$^a$, and A.~Yu.~Smirnov$^{a,b,c}$\\

\vspace{0.3cm}   

{\it $^a$ Department of physics, Tokyo Metropolitan University,  
Hachioji, Tokyo 192-0397, Japan} \\
{\it $^b$ ICTP, Strada Costiera 11, 34014 Trieste, Italy} \\
{\it $^c$ Institute for Nuclear Research of Russian Academy of Sciences,
Moscow 117312, Russia}

\end{center}

\begin{abstract}

As a result of identification of the solution to the solar 
neutrino problem, a rather precise relation 
$\theta_{sun} + \theta_C = \pi/4$ between the leptonic 1-2 
mixing angle $\theta_{sun}$ and the Cabibbo angle has emerged. 
It would mean that the lepton and the quark mixing angles 
add up to the maximal, suggesting a deep structure by 
which quarks and leptons are interrelated. We refer the relation 
``quark-lepton complementarity'' (QLC) in this paper.
We formulate general conditions under which the QLC relation is 
realized. We then present several scenarios which lead to the 
relation and elaborate on phenomenological consequences which 
can be tested by the future experiments. We also discuss 
implications of the QLC relation for the quark-lepton symmetry 
and the mechanism of neutrino mass generation.

\end{abstract}

%\pagebreak

%%%%%%%%%%%%%%%%%%%%%%%%%%%%%%%%%%%%%%%%%%%%%%%%%%%%%
\section{Introduction}
%%%%%%%%%%%%%%%%%%%%%%%%%%%%%%%%%%%%%%%%%%%%%%%%%

The most distinct feature of the lepton flavor mixing  
is the existence of two large mixing angles in the 
Maki-Nakagawa-Sakata (MNS) matrix \cite{MNS}, which is in 
sharp contrast to the CKM quark mixing \cite{CKM}.
One of the large angles comes from the atmospheric neutrino 
experiments \cite{SKatm} which have discovered the neutrino 
oscillation \cite{MNS,pontecorvo}, whereas the other one - from 
the solar \cite{solar} and the reactor neutrino observations 
\cite{KamLAND}. 
The atmospheric mixing is suspected to be 
maximal or close to the maximal, though the experiment gives only a 
mild constraint $36^{\circ} \leq \theta_{23} \leq 54^{\circ}$ \cite{hayato}.
On the other hand, the solar angle $\theta_{12}$ is known to be away 
from the maximal mixing value~\cite{SNO,SK_dn}.

It has been marked long time ago that the large mixing angle 
required for a solution of the solar neutrino problem 
may appear as a difference between the maximal mixing angle 
$\pi/4$ and the Cabibbo angle $\theta_C$, so that   
\be
\theta_{sun} + \theta_C = \frac{\pi}{4}, 
\label{equality}
\ee
or $\tan 2\theta_{sun} = 1/\tan 2\theta_C$~\cite{PS}. 
The equality holds with rather high accuracy 
as became clear by accumulating data of the solar neutrino experiments 
\cite{mirab}.  
Indeed, the global fit of the solar neutrino and KamLAND results   
gives~\cite{SNO,SK_dn,holanda,global}
\be
\theta_{sun} = 32.3^{\circ}~ \pm 2.4^{\circ}~~~~~(1\sigma).
\label{S_angle}
\ee
Taking the Cabibbo angle  at the $Z^{0}$ pole   
\be
\theta_C = 12.8^{\circ} \pm 0.15^{\circ} 
\label{C_angle}
\ee
%\cite{PDG}   
we obtain
\begin{equation}
\theta_{sun} + \theta_C = 45.1^{\circ} \pm 2.4^{\circ}~~~~~  (1\sigma) .
\label{sumdata}
\end{equation}

In terms of the oscillation observable the relation 
can be expressed as  
%observable
\be
\sin^2 \left( \frac{\pi}{4}  - \theta_C \right) = 0.284 \pm 0.002, 
~~~~\sin^2\theta_{sun} = 0.286 \pm 0.038 , 
%~~~~\sin^2\theta_{sun} = 0.289 \pm 0.041, 
\ee
so that 
\be
\Delta  \sin^2\theta_{12} \equiv 
\sin^2\theta_{sun} -
\sin^2 \left( \frac{\pi}{4} - \theta_C \right)
%= 0.005 \pm 0.043.  
= 0.002 \pm 0.040.  
%~~^{+ 0.040}_{-0.036}. 
\label{compare}
\ee
The deviation of the central value is well within the present 
experimental errors at $1\sigma$ CL. 
Notice that the best fit values of the solar angle from analyses 
of different groups have very small spread: 
$\theta_{sun} = 32.0^{\circ} - 33.2^{\circ}$. This shows stability 
of the result and may indicate that true value of $\theta_{sun}$ 
is indeed in this narrow interval, unless some systematic shift in 
the experimental data will be found. 
With this interval we obtain for the sum of the best fit angles 
\be
\theta_{sun} + \theta_C = 44.8^{\circ} - 46.0^{\circ}.
\ee

The equality (\ref{equality}) relates the 1-2 mixing angles 
in quark and lepton sectors, and if not accidental, 
implies certain relation between quarks and leptons.
It is very suggestive of a bigger structure in which quarks and 
leptons are complementary. 
The equality probably means a quark-lepton symmetry 
or quark-lepton unification~\cite{pati-salam} in some form. 
It may be considered as an  evidence of the 
grand unification, and/or certain flavor symmetry~\cite{raidal}. 
If not accidental, it can give a clue to 
understand the fermion masses in general context.   
In what follows we will call the equality (\ref{equality}) 
the quark-lepton complementarity (QLC) relation.

In this paper, 
we try to answer the following questions: 
Can the QLC relation be not accidental? 
What are the general conditions for the QLC relation? 
What is the underlying physical structure and the resultant 
scenarios that satisfy the conditions?
What are the experimental predictions of these scenarios and 
how can they be tested?
As a whole, we  explore experimental consequences and theoretical 
implications of the QLC relation.\\

The paper in organized as follows. In sec.~2 we formulate 
general conditions for the QLC relation.  
In sec.~3 and 4 we elaborate on various scenarios 
which realize the relation (\ref{equality}).  
In sec.~3 a possibility of ``bimaximal minus CKM mixing'' is studied.  
In sec.~4 we consider single maximal mixing scenarios. 
In sec.~5 the predictions by various scenarios are summarized.
In sec.~6 we give a summary with brief comment on 
how to test them experimentally. Some theoretical implications of 
the QLC relation and heuristic remarks are also presented.

In secs.~3 and 4 we give detailed and comprehensive description of 
possible phenomenological scenarios providing for each case with 
comments on implications for neutrino mass matrix and 
quark-lepton symmetry. For those who want to avoid these details 
we recommend, after reading sec.~2, to go directly to sec.~5 in which 
an overview of phenomenological aspects of our results are summarized, 
in particular in Table 1.
One can go back for details of particular scenarios to secs.~3 and 4.

%%%%%%%%%%%%%%%%%%%%%%%%%%%%%%%%%%%%%%%%%%%%%%%%%%%%%%%%%%%%%%%
\section{General conditions for the quark-lepton complementarity relation}
%%%%%%%%%%%%%%%%%%%%%%%%%%%%%%%%%%%%%%%%%%%%%%%%%%%%%%%%%%%%%%%%%

The lepton mixing matrix $U_{MNS}$ is defined as 
\be
U_{MNS} = U^{\dagger}_e U_{\nu}~, 
\label{m-mns}
\ee
where  $U_e$  and   $U_{\nu}$ are the transformations of the left 
handed components which diagonalize the mass matrices of the 
charged leptons and neutrinos respectively. 
In the standard parameterization \cite{PDG} 
the MNS matrix reads\footnote{
%%%%%%%%%%%%%%%%%%%%%% footnote %%%%%%%%%%%%%%%%%%%%%%%%%
While the form in (\ref{param}) utilizes a slightly non-standard way 
of introducing a CP violating phase into the MNS matrix \cite{chau-keung}, 
it can be shown that the correspondence of the angles with 
the experimental observable is the same as those of the standard 
parameterization \cite{PDG}. 
}
\be
U_{MNS} = R_{23} \Gamma_{\delta_l} R_{13} R_{12},  
\label{param}
\ee
where  $R_{ij}$ is the  matrix of rotation in the $ij$ - plane.  
In this form, the angle of 1-2 rotation is identified  with the solar angle, 
$\theta_{12} = \theta_{sun}$, the angle of 2-3 rotation - with 
the atmospheric angle, $\theta_{23} = \theta_{atm}$,  and 
$\theta_{13}$ - with the angle  restricted by the  
CHOOZ experiment~\cite{CHOOZ}. 
The matrix with the CP-violating phase is parameterized as    
$$
\Gamma_\delta \equiv diag (1, 1, e^{i\delta_l}).  
$$ 
%where $\delta$ is the CP violation phase. 
%
To identify the mixing angles with those 
measured in experiments one should reduce a given mixing 
matrix to the form (\ref{param}). 

Let us formulate general conditions which lead to the QLC relation.

%%%%%%%%%%%%%%%%%%%%%%%%%%%%%%%%%%%%%%%%%%%%%%%%%%%%%%
\subsection{Single maximal or bi-maximal}
%%%%%%%%%%%%%%%%%%%%%%%%%%%%%%%%%%%%%%%%%%%%%%%%%%%%%%%%

In principle, it is enough to have a single maximal mixing, 
that is $R_{12}^m \equiv R_{12}(\pi/4)$, to realize relation 
(\ref{equality}).  
However, existence of maximal or near maximal 
2-3 leptonic mixing hints that 
whole pattern of fermion mixings may be generated as 
a combination of no mixing, a maximal and the CKM mixings. 
Namely, we can speak on the scenario characterized by 
\be
\mbox{``bi-maximal minus CKM mixing''.}
\label{scenario}
\ee
Because it is very predictive and the easiest 
to test experimentally, it deserves a separate description from 
more general cases.
A possibility of the lepton mixing as small 
deviation from the bi-maximal mixing~\cite{bm} has been extensively 
discussed recently~\cite{bimaximal} but without 
identification of small deviation with the quark mixing. 
See, however, the first reference in ~\cite{bimaximal}.
Relation (\ref{equality}) allows to restore the bi-maximal mixing 
\cite{bm} as the element of underlying theory \cite{raidal}.

It should be stressed \cite{concha-smir1} 
that the present data do not yet give strong bound on 
deviation of 2-3 mixing from the 
maximal, which can be characterized by 
\begin{equation}
D_{23} \equiv 0.5 - \sin^2 \theta_{23}. 
\label{devia}
\end{equation}
It is constrained by $|D_{23}| \leq 0.16$, 
or $|D_{23}|/\sin^2 \theta_{23} \leq 0.47$ 
at 90\% CL \cite{hayato}. 
Furthermore, the latest analysis, 
(without renormalization of the original fluxes) 
shows some excess of the $e$-like events at sub-GeV energies 
and the absence of excess in the multi-GeV sample, 
thus giving a hint to non-zero $D_{23}$ \cite{concha-smir2}.

In the  scenario (\ref{scenario}), 
one expects the deviation to be  small:  
$\pi/4 - \theta_{23} \lsim \theta^{CKM}_{23}$, or   
\be
|D_{23}| \lsim \sin \theta_{23}^{CKM} \approx  
V_{cb}   \simeq \sin^2\theta_C \simeq 0.04.  
\label{devia-gene-l}
\ee
For specific scenarios  see sec.~3. 
The next generation long-baseline experiments, in particular the 
JPARC-SK, will be sensitive to $|D_{23}| \sim 0.05$ 
\cite{JPARC,MSS,AHKSW}.
Also it would be a challenge for the future atmospheric neutrino 
experiments to achieve the required sensitivity. 
Establishing the deviation from the maximal mixing more significant 
than the one in (\ref{devia-gene-l}) 
%and (\ref{devia-gene-nu}) 
will exclude the scenario (\ref{scenario}).

If the bi-maximal scenario is not realized and $D_{23}$ is large, 
an additional 1-3 rotation (apart from 1-3 CKM rotation) should 
be considered. 
Indeed, generically, the same symmetry ({\it e.g.}, $Z_2$) leads 
to the maximal 2-3 mixing and simultaneously vanishing 1-3 
mixing~\cite{grimus}. Therefore, the deviation from maximal 2-3 angle,   
$D_{23}$, which implies violation of the symmetry,  
should also be accompanied by a non-zero 1-3 mixing. In this case, 
predictability will be lost unless one imposes the condition 
that such an additional 1-3 rotation is very small.

%%%%%%%%%%%%%%%%%%%%%%%%%%%%%%%%%%%%%%%%%%%%%%%%%
\subsection{Order of rotations} 
%%%%%%%%%%%%%%%%%%%%%%%%%%%%%%%%%%%%%%%%%%%%%%%%%

To reproduce the equality (\ref{equality}) 
exactly one needs to have the following order of rotations:  
\be 
U_{MNS} = \cdot \cdot \cdot
R_{23}^m \cdot \cdot \cdot
R_{12}^{CKM \dagger} R_{12}^m, ~~~~
{\rm or }~~~~
U_{MNS} = \cdot \cdot \cdot 
R_{23}^m \cdot \cdot \cdot 
R_{12}^m  R_{12}^{CKM \dagger}.   
\label{gen1}
\ee
That is, the maximal and the CKM rotations must be attached with each other. 
Here, $R_{ij}^{CKM} \equiv R_{ij}(\theta^{CKM}_{ij})$ 
describes the CKM rotation  in the $ij$-plane, and 
$R_{ij}^m$ denotes the maximal mixing rotations, 
$R_{ij}^m \equiv R_{ij}(\pi/4)$.
In (\ref{gen1}) $``\cdot \cdot \cdot''$ denotes possible 
insertion of the CKM rotations, $R_{23}^{CKM}$ and $R_{13}^{CKM}$.  
(The similar structure holds also in the case that $R_{23}$ 
is not maximal.)
The complete CKM matrix is parametrized as  
\be
V^{CKM} =  R_{23}^{CKM} \Gamma_{\delta_{q}} 
R_{13}^{CKM} R_{12}^{CKM}. 
\label{CKM}
\ee

The reversed ordering of maximal mixing rotations in 
(\ref{gen1}), namely $R_{12}^m \cdot \cdot \cdot R_{23}^m$,  
would lead to an unacceptably large 1-3 mixing:  
$\sin \theta_{13} = 0.5$  and incorrect 1-2  mixing, 
$\theta_{sun} \sim \pi/6 \pm \theta_C$,
after reducing the mixing matrix to the form  (\ref{param}).   

Two other CKM rotations, $R_{23}^{CKM}$ and $R_{13}^{CKM}$, 
can be located in any place indicated by dots. 
%The effect of CKM rotations $R_{23}^{CKM}$ and $R_{13}^{CKM}$  
Their effect on the relation  (\ref{equality}) is negligible even if  
they  are situated in the right-hand side of the combinations in 
(\ref{gen1}) or between two 1-2 rotations. 
The largest possible deviation appears for the case 
$R_{12}^m R_{12}^{CKM \dagger} R_{23}^{CKM}$ which, however, 
reduces to a small unobservable  correction: 
\be
\sin^2\theta_{sun} \rightarrow \sin^2\theta_{sun} (1 - V_{cb}^2), 
\ee 
where $\sin \theta_{23}^{CKM} \approx V_{cb} = 0.04$ 
($\theta_{23}^{CKM} = 2.3^{\circ}$). 
In what follows we will neglect these type of corrections to the 1-2 
mixing. However,  
position of small CKM rotations can become important for other 
observable such as $U_{e3}$ or deviation of the 2-3 mixing 
from the maximal one.  

We will also consider the combination  
\be
U_{MNS} =  \cdot \cdot \cdot 
R_{12}^{CKM \dagger} R_{23}^m \cdot \cdot \cdot  
R_{12}^m  
\label{gen3}
\ee
which is not excluded experimentally, 
though leading to the QLC relation (\ref{equality}) 
only in an approximate way.

%%%%%%%%%%%%%%%%%%%%%%%%%%%%%%%%%%%%%%%%%%%%%%%%%%%%%%%%
\subsection{CKM matrix and the quark-lepton symmetry}
%%%%%%%%%%%%%%%%%%%%%%%%%%%%%%%%%%%%%%%%%%%%%%%%%%%%%%%%

The natural framework in which the CKM angles appear  in  
the lepton mixing is the quark-lepton symmetry \cite{pati-salam} according 
to which in a certain basis 
\be 
V_{\nu} = V_{u} = V^{CKM \dagger}~~ {\rm or}~~ V_{l} = V_{d} = V^{CKM}.  
\label{gu}
\ee
Then according to the definition (\ref{m-mns}) 
in both cases the CKM matrix will appear in the leptonic matrix 
as hermitian conjugate,  
\be 
U_{MNS} \propto \cdot \cdot \cdot  
V^{CKM \dagger} \cdot \cdot \cdot = \cdot \cdot \cdot  
R_{12}^{CKM \dagger} R_{13}^{CKM \dagger} 
R_{23}^{CKM \dagger} \cdot \cdot \cdot . 
\ee 
Therefore, some permutations of $R_{12}^{CKM \dagger}$ and 
other matrices
%, $R_{23}^{CKM \dagger}$ (lepton scenario, subsec. 3.2), 
%or $R_{23}^{CKM \dagger}R_{23}^m$ (neutrino scenario, subsec. 3.1),  
are necessary which lead to a violation of the 
exact relation (\ref{equality}). The smallest corrections are 
produced  when only $R_{12}^{m}$ appears right next to 
$V^{CKM \dagger}$ on the RHS of the mixing matrix (\ref{gen1}).
In this case  
$\Delta \sin^2\theta_{12} \sim \sin\theta_C V_{cb}^2$. 
%%
%In the former case the correction is very small, 
%$\sim \sin\theta_C V_{cb}^2$, while in the latter it is 
%sizable, $\sim \sin\theta_C$.
%%

It is possible that the quark-lepton connection is not realized in 
a straightforward way as in (\ref{gu}). The Cabibbo angle 
could be the universal parameter which controls the whole structure 
of fermion masses and therefore appears in many places such as 
mass ratios and mixing parameters (see  sec. 6).

%%%%%%%%%%%%%%%%%%%%%%%%%%%%%%%%%%%%%%%%%%%%%%%%%%%%%%%%
\subsection{Naturalness}
%%%%%%%%%%%%%%%%%%%%%%%%%%%%%%%%%%%%%%%%%%%%%%%%%%%%%%%%

In underlying models one expects that some deviation from the
exact QLC relation always exists. It can be parametrized as
\be
\theta_{sun} - \frac{\pi}{4} + \theta_C = \Delta \theta_{12}(X_i),
\ee
where $X_{i}$ denote  parameters of a model. Note that 
$\Delta \sin^2 \theta_{12} = \sin 2 \theta_{sun} \Delta \theta_{12}$.
Then, one should require that $\Delta \theta_{12}(X_i)$ 
is very small in whole allowed ranges of the parameters $X_i$. 
Otherwise, the QLC relation appears as a result of fine 
tuning of several parameters and in this sense 
turns out to be {\it unnatural} or  {\it accidental}.

This leads to  immediate and non-trivial conditions:
$\Delta \theta_{12}(X_i)$
should not depend on the masses of quarks and leptons or 
the dependence must be weak. Indeed, masses of down quarks 
and charged leptons for the first and the second generations 
(which are relevant here) are substantially different. 
Therefore, one would not expect an appearance of the
same mixing angle $\theta_C$ in the quark and the lepton sector.
The quark-lepton symmetry should be realized in terms of mixings and
not masses.

%%%%%%%%%%%%%%%%%%%%%%%%%%%%%%%%%%%%%%%%%%%%%%%%
\subsection{Effect of  CP Violation} 
%%%%%%%%%%%%%%%%%%%%%%%%%%%%%%%%%%%%%%%%%%%%%%%%%%%%%%%

Diagonalization of the neutrino and charge lepton mass matrices can 
lead to the CP-violating phases in $U_{l}$ and $U_{\nu}$ 
(which eventually will be reduced to the unique phase 
$\delta_l$ in $U_{MNS}$). 
This can be described by the phase matrices 
$$
\Gamma_{\delta' \delta} =  diag (e^{i\delta'}, 1 , e^{i\delta})
$$  
which appear in various places of the products 
(\ref{gen1}). To keep the equality 
(\ref{equality}), the matrices $\Gamma_{\delta, \delta'}$ should not 
be between  $R_{12}^{CKM}$ and $R_{12}^m$, or the 
corresponding phases  should be small enough. 
Indeed,  the structure $R_{12}^m \Gamma_{\delta' 0} R_{12}^{CKM}$ leads 
to 
\be
\Delta \sin^2 \theta_{12} = 
\frac{1}{2} \sin 2\theta_C  (1 - \cos \delta').
\ee
We find that the QLC-relation (\ref{equality}) 
is satisfied within $1\sigma$, provided that     
$\delta' < 34^{\circ}$.

With the additional phase $\delta'$, the QLC relation 
(\ref{equality}) appears as a result of fine tuning of 
the parameters and therefore is not natural. 
Hence, we restrict ourselves into the choice 
$\Gamma_{\delta} \equiv  diag(1, 1, e^{i \delta})$ 
in the rest of the paper.  
Then, the place where we can insert the phase matrix is unique: 
it can be easily checked that  all other possible insertions either 
can be reduced to this possibility or 
lead to zero CP-violation.  

Furthermore, the $\delta$ dependence comes into expressions of 
the various mixing matrix elements and the Jarlskog 
invariant  only together with 
$|V_{cb}| \simeq 0.04$.
Indeed,  in the limit of zero rotation $R_{23}^{CKM} = 1$ 
(and $R_{13}^{CKM} = 1$)  the mixing matrices $U_{MNS}$ 
(\ref{gen1})  (\ref{gen3}) are reduced to 
\be
R_{23}^m R_{12}^m R_{12}^{CKM \dagger} ~{\rm or} ~~
R_{12}^{CKM \dagger}R_{23}^m R_{12}^m.
\ee
In both cases any insertions of the 
phase matrices $\Gamma_\delta$ will not lead to physical CP violation 
phase. Therefore, in the limit $V_{ub} = 0$ 
the CP-violation effects (Jarlskog invariant) 
are proportional to $V_{cb}$:  
\be
J_{lep} \equiv Im \left[
U_{\alpha i} U^*_{\beta i}  U^*_{\alpha j} U_{\beta j}
\right]  \propto 
V_{cb}. 
\ee

We note, in passing that if $V^{CKM}$ is the only origin 
of the CP violation, namely, if $\delta=0$,  
we obtain generically 
\be
\sin \delta_l = \frac{V_{ub}}{U_{e3}} \sin \delta_{q}, 
\ee  
where $\delta_{q}$ is the phase in the CKM matrix. 
Since $U_{e3}$ can be larger than $V_{ub}$ due to 
contribution induced by ``permutations'', 
the leptonic CP violation phase is strongly 
suppressed in this case. 
Induced CP violation associated with $\delta$ can be much larger.

%%%%%%%%%%%%%%%%%%%%%%%%%%%%%%%%%%%%%%%%%%%%%%%%%%%%%%%%%%%%
\subsection{Renormalization group effect} 
%%%%%%%%%%%%%%%%%%%%%%%%%%%%%%%%%%%%%%%%%%%%%%%%%%%%%%%%%%%%%

The QLC relation (\ref{equality}) holds at low energies. 
However, the quark-lepton symmetry (unification) which leads to 
(\ref{equality}) is realized most probably at some high energy scales, 
{\it e.g.}, the grand unification scale. 
To guarantee the QLC relation at high energies one should 
require that the renormalization group effects on the equality  
from this high scale to the low energy scale are small. 
In the Standard Model (SM), or 
Minimal Supersymmetric Standard Model (MSSM) 
the renormalization of the Cabibbo angle is indeed small. 
For instance, in MSSM with $\tan\beta =  50$
the  parameter  $\sin \theta_C$ decreases from 0.2225 at the 
$m_Z$ down to 0.2224 at the $10^{16}$  GeV~\cite{C-running}.

The renormalization effect on the leptonic $\theta_{12}$ 
depends on the type of mass spectrum of light neutrinos. 
For the  spectrum with normal mass hierarchy, 
$m_1 < m_2 \ll m_3$, the effect is negligible. 
In contrast, in the case of quasi-degenerate spectrum, 
$m_1 \approx  m_2 \approx m_3 = m_0$,  or the spectrum with
inverted mass hierarchy the effects can be large~\cite{renorm}.

In the limit of small 1-3 mixing $\theta_{13} \ll 10^{\circ}$, 
the running is determined by~\cite{lindner}
\be
\frac{d\theta_{12}}{dt} \approx - \frac{C y_{\tau}^2}{32 \pi^2} \sin
2\theta_{12}
\sin^2 \theta_{23} \frac{|m_1 e^{i\phi_1} +  m_2 e^{i\phi_2 }|^2}{\Delta
m^2_{sun}}, 
\label{ren}
\ee
where $t \equiv ln(\mu/\mu_0)$, $\mu$ is the renormalization scale, $C = 1$ in
the MSSM and
$C = - 3/2$ in the SM;
$y_{\tau}$ is the Yukawa coupling of the tau lepton:
\be
\frac{C y_{\tau}^2}{32 \pi^2} \approx
\left\{
\begin{array}{ll}
0.3 \cdot 10^{-6}, & \hspace{0.5cm} \mbox{SM} \\
0.3 \cdot 10^{-6} (1 + \tan^2 \beta),  & \hspace{0.5cm} \mbox{MSSM} 
\end{array}
\right.
\ee
and $\tan \beta$ is the usual ratio of the VEV's.
In Eq. (\ref{ren})
$\phi_1$ and  $\phi_2$ are the Majorana phases of the eigenstates $\nu_1$ 
and
$\nu_2$.  According to (\ref{ren}), the running effect is proportional
to the absolute mass scale squared and the relative phase difference:
$\dot{\theta}_{12} \sim m_0^2 \cos(\phi_2 - \phi_1)/2$.
In SM and in MSSM with $\tan \beta < 10$ the corrections are small 
even for quasi-degenerate mass spectrum. 
In MSSM with large $\tan \beta$ ($\tan \beta = 50$) 
one finds that $\Delta \theta_{12} \sim \theta_{12}$
even for the common scale $m_0 \sim 0.1$ eV \cite{lindner} 
as a result of running from the scale of the RH neutrinos  ($10^{10} -
10^{12}$ GeV) or the GUT scale.
Clearly, such a large correction destroys the QLC relation, 
which leads us to the following conclusions:

\vspace{0.2cm}
\noindent
1). The QLC relation is not violated by the renormalization effect 
in the SM and in MSSM with small $\tan\beta$ even for the 
quasi-degenerate mass spectum of neutrinos.

\vspace{0.2cm}
\noindent
2). In MSSM with large $\tan \beta$ and the quasi-degenerate mass 
spectrum the corrections are in general large. 
Furthermore, the corrections depend on other continuous 
(and presently unknown) parameters: $\phi_i$, $m_0$
(and also $\theta_{13}$), so that the QLC relation would 
require fine tuning of several parameters.  Therefore, 
the QLC relation, once it is established with a good accuracy, 
testifies against such models, unless the required tuning is 
a natural outcome of an additional symmetry.
Notice that according to (\ref{ren}),  the corrections can be 
strongly suppressed if the quasi-degenerate mass eigenstates
$\nu_1$ and $\nu_2$ have opposite CP parities: 
$\phi_2 - \phi_1 \approx \pi$ \cite{renorm}.

\vspace{0.2cm}
\noindent
3). In some cases the renormalization effect can help to reproduce 
the QLC relation (see sec. 3.1).

%%%%%%%%%%%%%%%%%%%%%%%%%%%%%%%%%%%%%%%%%%%%%%%%%%%%
\subsection{Basis dependence}
%%%%%%%%%%%%%%%%%%%%%%%%%%%%%%%%%%%%%%%%%%%%%%%%%%%%%%

%The flavor mixing matrix consists of the rotations that arise 
%as a result of diagonalization of the mass matrices. 
The form of the mass matrices and diagonalizing rotations 
depend on basis of the quark and lepton states. 
Let us introduce a basis called the symmetry basis by which a 
symmetry that determines the structure of mass matrices is 
defined. (In some publications this basis is named as the 
Lagrangian basis.)

In the symmetry basis, both the neutrino and the charged fermion 
mass matrices, in general, are not diagonal and therefore 
both produce rotations which make up the MNS matrix. 
In what follows we will consider several realizations of the 
structure of lepton mixing matrix, (\ref{gen1}) and (\ref{gen3}). 
They differ by the origin of the large (maximal) angle 
rotations:  the neutrino or the  charge lepton sectors.
These different realizations have different theoretical and 
experimental implications.

%%%%%%%%%%%%%%%%%%%%%%%%%%%%%%%%%%%%%%%%%%%%%%%%%%%%%%%%%%%
\section{Bi-maximal minus  CKM mixing}
%%%%%%%%%%%%%%%%%%%%%%%%%%%%%%%%%%%%%%%%%%%%%%%%%%%%%%%%%%%%

In this section we will consider different realizations of the 
possibility (\ref{scenario}) in which only maximal mixings 
and the CKM rotations are involved in formation of the 
fermion  mixing matrices.

%%%%%%%%%%%%%%%%%%%%%%%%%%%%%%%%%%%%%%%%%%%%%%%%%%%%%%%%%%%%%%%
\subsection{Bi-maximal mixing from neutrinos}    
%%%%%%%%%%%%%%%%%%%%%%%%%%%%%%%%%%%%%%%%%%%%%%%%%%%%%%%%%%%%%%%%%

Let us assume that in the symmetry basis the bi-maximal mixing 
originates from the neutrino mass matrix,  whereas the charged 
lepton mixing matrix coincides with the CKM matrix: 
\be 
U_{\nu} = R_{23}^{m} R_{12}^m, ~~~~ U_l  =  V^{CKM}. 
\label{lep-nusc}
\ee 
Then the lepton mixing matrix equals 
\be 
U_{MNS}  = V^{CKM \dagger}\Gamma_\delta  R_{23}^m R_{12}^m =
R_{12}^{CKM \dagger} R_{13}^{CKM \dagger} R_{23}^{CKM \dagger} 
\Gamma_{\delta} 
R_{23}^m R_{12}^m , 
\label{mns-nu1}
\ee
where we have introduced the phase matrix $\Gamma_{\delta}$ 
following our general prescription described in Sec. 2.

In the quark sector we have  
\be
V_u = I, ~~~~~V_d = V^{CKM},  
\ee
so that the second equality in (\ref{lep-nusc}) 
implies the quark-lepton symmetry relation,  $V_l = V_d$. 
We also assume that the neutrino Dirac matrix is diagonal due to 
the equality 
\be
m_{\nu}^D = m_u.
\label{up-nu}
\ee
Then, the bi-maximal rotation of neutrinos follows from
the seesaw mechanism \cite{seesaw} and the specific  structure 
of the mass matrix of right-handed (RH) neutrinos. 
Notice that the bi-maximal mixing can be related to the 
quasi-degenerate type mass spectrum of neutrinos.  
Such a possibility for the bi-maximal neutrino mixing  and 
general matrix $U_l$, not necessarily 
related to $V^{CKM}$, has been discussed recently in 
\cite{bimaximal}.

The problem in this scenario is that in spite of the equality 
$V_d = V_l$ the mass eigenvalues are different: 
$m^{diag}_d \neq m^{diag}_l$, where 
$m^{diag}_l \equiv diag (m_e, m_{\mu}, m_{\tau})$.  
Therefore, the mass matrices are also different.  
Some special conditions have to be met for the matrices 
such that they produce the same mixing despite the different 
eigenvalues. A possibility is  the singular mass 
matrices for which different (strong) mass hierarchies 
can be reconciled with  approximate  equality of the  
of mixing matrices~\cite{singular}. 

Let us discuss the phenomenological consequences of this scenario.  

\vspace{0.2cm}   

\noindent
1). The  mixing matrix (\ref{mns-nu1}) does not satisfy the conditions
(\ref{gen1}) and therefore the relation (\ref{equality}) receives 
corrections 
\be
\sin \theta_{sun} = \sin \left(\frac{\pi}{4} -\theta_C  \right) +
\frac{\sin \theta_C}{2} (\sqrt{2} - 1 - V_{cb}\cos \delta).
\label{solangle}
\ee
Numerically, 
we obtain for $\theta_{sun}$  
\be
\theta_{sun} = 35.4^{\circ} \pm 0.3^{\circ}, ~~~~ 
\sin^2 \theta_{sun} = 0.335 \pm 0.005, 
\label{sun-nu1}
\ee
and for the deviation parameter 
\be 
\Delta \sin^2 \theta_{12} \approx  
\sin \theta_{sun} \sin\theta_C (\sqrt{2} - 1 - |V_{cb}|\cos \delta)
=  0.046 - 0.056, 
\label{SC_deviat-nu1}
\ee
where the intervals indicate uncertainty due to the unknown phase 
$\delta$. The deviation in (\ref{SC_deviat-nu1}) is $15-20$ \%. 
%where the numbers are calculated from (\ref{sun-nu1}). 
It corresponds  to 
$ 
\theta_{sun} + \theta_C - \frac{\pi}{4} \simeq  2.9^{\circ} -  3.6^{\circ}. 
$
%or, the central value of angles are given by 
%$\theta_{sun} + \theta_C = 48.3^{\circ}$. 
%
Therefore, one needs to measure $\sin^2 \theta_{sun}$ with 
better than 10\% accuracy to establish this difference. 
According to the estimations given in \cite{bahcall-pena}, the  
future solar neutrino and the KamLAND experiments may have a 
sensitivity of $\simeq 4$ \% to $\sin^2 \theta_{sun}$, 
provided that $\theta_{13}$ is measured, or severely restricted. 
The sensitivity of a dedicated reactor $\theta_{12}$ experiment 
can reach $\simeq 3$ \% \cite{MNTZ}. The errors quoted are at 
the confidence level of 1 $\sigma$. 
So with such an accuracy the equality (\ref{solangle}) can be 
established at about $(4 - 5) \sigma$.

\vspace{0.2cm}   

\noindent
2). For 1-3 mixing we obtain
\be
\sin \theta_{13} = - \frac{1}{\sqrt{2}} 
\sin \theta_C  
\left(1 - |V_{cb}| \cos \delta \right) + V_{ub}, 
\label{13-indu}
\ee
where the first dominant term is induced by permutation of the 
Cabibbo rotation $R_{12}^{CKM}$ with the nearly maximal 2-3 rotation.
%and the second term can be neglected. 

The two elements of $U_{MNS}$, $|U_{e3}|$ and $|U_{\mu 3}|$, 
are connected by a simple relation  
\be 
|U_{e3}|^2 = \tan^2 \theta_{C} |U_{\mu 3}|^2 
\label{e3-mu3}
\ee
which does not depend on $\delta$ and  
$\theta^{\nu}_{23}$  
(the latter is taken to be $\pi/4$ in this section), 
and represents the characteristic feature of the scenario 
of bi-large mixing from neutrinos (see sec.~4). 
Using the Super-Kamiokande bound \cite{hayato}
$0.34 \leq |U_{\mu 3}|^2 \leq 0.66$, we obtain the prediction 
for $|U_{e3}|^2$: 
\be 
\sin^2 \theta_{13} = 0.026 \pm 0.008
%0.0177 \leq |U_{e3}|^2 \leq 0.0345, ~~~ or ~~~
%0.075 \leq \sin^2 2\theta_{13} \leq 0.11
\label{e3-mu3_2}
\ee
which is just below the CHOOZ bound 
and falls into the region of sensitivity of the  
next generation accelerator \cite{JPARC,MINOS,OPERA,NuMI,SPL} 
and the reactor  experiments \cite{MSYIS,reactor_white}.

\vspace{0.2cm}   

\noindent
3). The deviation of 2-3 mixing from the maximal can be written as 
\be
D_{23} = 
%\frac{1}{2} \cos 2\theta^{\nu}_{23} + 
\frac{1}{2} \sin^2 \theta_C 
+ \cos^2 \theta_C |V_{cb}| \cos \delta,  
\label{d23_bl_nu}
\ee
where the two terms are of the same order.  
Numerically it gives 
\be
D_{23} =  0.025 \pm 0.039, 
\ee
and the interval is due to  the unknown  
CP violating phase. Maximal possible value of $D_{23}$ is at 
the level of sensitivity of the J-PARC experiment \cite{JPARC}.

\vspace{0.2cm}   

\noindent
4). For the leptonic Jarlskog invariant we obtain  
\be
J_{lep} =  \frac{1}{8 \sqrt{2}} \sin{2\theta_{C}} 
|V_{cb}| \sin{\delta}
\simeq 
1.5 \times 10^{-3} \sin{\delta}. 
\label{J-nu1}
\ee
It is a factor of $\simeq$ 30 smaller than the maximal
value of $J_{lep}$ allowed by the CHOOZ constraint: 
\be
J^{max}_{lep} \simeq 0.04 \sin{\delta}. 
\label{jmax}
\ee
We note that $J_{lep}$ vanishes in the two-flavor limit 
$\theta_{13} \rightarrow 0$, as it should, because the limit 
implies $\theta_{C} \rightarrow 0$ (ignoring $V_{ub}$), 
as one can see from (\ref{13-indu}).

%Though consistent, 
The smallness of $J_{lep}$ in (\ref{J-nu1}) 
despite the relatively large $\sin \theta_{13}$ means that 
the way of introducing the CP violating phase $\delta$ in 
(\ref{mns-nu1}) is not quite general. As we have shown in sec. 2.4
the induced part  is proportional to $V_{cb}$ and 
if the CKM matrix is the only source of 
CP violation the resultant leptonic CP violation is 
extremely small.\\

Let us consider a possibility that the value of $\theta_{12}$ 
given in (\ref{sun-nu1}) is realized  
at high-energy scale, and it  diminishes when   
running from high to low energy scales. So the better 
agreement with the QLC relation is achieved at the electroweak scale.
As we have discussed in sec. 2.5,
a substantial effect due to renormalization can be obtained 
in the MSSM with large $\tan \beta$ and quasi-degenerate 
neutrino mass spectrum. In this case, however, 
running toward low energies leads to an increase of 
$\theta_{12}$, as follows from (\ref{ren}) for negligible 
$\sin \theta_{13}$.
Therefore, to diminish  $\theta_{12}$, one needs
(i) to suppress the main term given in (\ref{ren}),  and 
(ii) to take into account the effect due to non-zero 1-3 mixing.
The former can be reached in the case of opposite CP-parities of
$\nu_1$ and $\nu_2$.
As far as the latter is concerned, it was shown in~\cite{lindner} 
that if $\phi_2 - \phi_1 \approx \pi$ the decrease of 
$\theta_{12}$ by $3^{\circ} - 5^{\circ}$ can be easily achieved by 
running down from $(10^{10} - 10^{13})$ GeV for 
$\theta_{13} = 5^{\circ} - 10^{\circ}$.

%%%%%%%%%%%%%%%%%%%%%%%%%%%%%%%%%%%%%%%%%%%%%%%%%%%%%
\subsection{Bi-maximal mixing from charged leptons}
%%%%%%%%%%%%%%%%%%%%%%%%%%%%%%%%%%%%%%%%%%%%%%%%%%%%%

Let us assume that the bi-maximal mixing appears 
from diagonalization of the charged lepton mass matrix,
%\cite{}, 
whereas the 
CKM rotation originates from the neutrino sector:  
\be
V_{\nu} = V^{CKM \dagger}, ~~~~~~ 
V_l =  R_{12}^{m \dagger}  R_{23}^{m  \dagger}.
\label{rot-l}
\ee
This possibility has been suggested in~\cite{raidal}. Our 
predictictions, however, differ from those obtained in ~\cite{raidal}.   

Notice that in $U_l$ the 1-2 and 2-3 rotations need to be 
permuted in comparison with the standard definition of the 
bi-maximal matrix to produce correct order of rotations in $U_{MNS}$.  
The lepton mixing matrix with the CP phase $\delta$ is given by 
\be
U_{MNS}  = R_{23}^m \Gamma_{\delta} R_{12}^m  V^{CKM \dagger} = 
R_{23}^m \Gamma_{\beta} 
R_{12} (\pi/4 - \theta_{12}^{CKM}) 
R_{13}^{CKM \dagger} 
R_{23}^{CKM \dagger}.  
\label{mns-l1}
\ee
In the quark sector we assume the left rotations  
\be
V_{u} =  V^{CKM \dagger}, ~~~~~ V_d = I. 
\label{quark-leps}
\ee
%which reproduce the CKM mixing matrix for quarks. 

The former  relations in (\ref{rot-l}) and (\ref{quark-leps})
imply the quark-lepton symmetry, 
$V_{\nu} = V_{u}$. This in turn can   
originate from the equality of the up-quark and
the neutrino Dirac mass matrices, 
$m_{u} = m^D_{\nu}$ as in (\ref{up-nu}), 
under the assumption (in the seesaw context) that the
Majorana mass matrix of the right handed neutrinos does
not produce any additional rotations \cite{raidal}.
However, the latter equalities in (\ref{rot-l}) and (\ref{quark-leps}) 
require a departure from the simple quark-lepton symmetry. 
They can be easily accommodated in the 
``lopsided'' schemes \cite{lopsided} of the SU(5) GUT. 
However, the relation (\ref{up-nu}) is not explained  
in SU(5). 
In SO(10) models which naturally lead to  (\ref{up-nu}), 
on the other hand, the lopsided scenario requires 
further complications.
The scenario does not appear to follow naturally from 
the grand unified models. 
Notice that the problem of equal mixings but different 
masses outlined in sec. 3.1 exists also here: In the basis where 
$m_d$ and $m_l$ are diagonal, that is $V_d = V_l = I$, the eigenvalues 
of mass matrices are different. In another words 
the question is why $m_d$ and $m_l$ are diagonal in the same basis.

Let us spell out the consequences of the lepton bi-maximal scenario.

\vspace{0.2cm}   

\noindent
1). The matrix (\ref{mns-l1}) reproduces the relation (\ref{equality})
almost exactly, 
\be 
\sin \theta_{sun} = 
\sin \left( \frac{\pi}{4} - \theta_C \right) - 
\frac{1}{2} 
\sin \theta_{sun} |V_{cb}|^2 - 
\cos \theta_{sun} |V_{cb}||V_{ub}|.
\label{sun-l1}
\ee
%
%where we have ignored the terms of the order $|V_{cb}|^4$.
Numerically we obtain 
\be
\Delta \sin^2 \theta_{12} = 
-\sin^2 \theta_{sun} |V_{cb}|^2  \simeq 
- 6 \times 10^{-4}
\ee
and $\Delta \theta_{12} = 0.04^{\circ}$.

\vspace{0.2cm}   

\noindent
2). For 1-3 mixing we have  
\be
\sin \theta_{13} =  - 
\sin \theta_{sun} |V_{cb}| - \cos \theta_{sun} |V_{ub}| 
\approx - \sin \theta_{sun} |V_{cb}|,
\label{13-lep}
\ee
where the induced (by the permutation of matrices) 
first term dominates. 
Eq. (\ref{13-lep})  leads to a very small value, 
$|U_{e3}|^2 \simeq 5 \times 10^{-4}$, or 
$\sin^2 2\theta_{13} =  1.9 \times 10^{-3}$ 
($\theta_{13}=1.2^{\circ}$). 
It is beyond reach of the proposed superbeam experiments and 
may be reached only by neutrino factory \cite{nufact}. 
We note that $U_{e3}$ being of the order 
$\lambda^2$ in the Wolfenstein parametrization \cite{Wparam}, our result 
(\ref{13-lep}) differs from the estimation made in \cite{raidal}. 

\vspace{0.2cm}   

\noindent
3). The 2-3 mixing angle is determined, ignoring the terms of 
the order $|V_{cb}|^2$, by 
\be
\sin \theta_{23} = 
\sin \left(\frac{\pi}{4} - \theta_{23}^{CKM}\right) +
\frac{1}{\sqrt{2}}
\left(1-\cos \theta_{sun}\cos \delta \right) |V_{cb}|.
\label{23-lep}
\ee
The second term in the RHS of (\ref{23-lep}) is small, 
and the relation
$\theta_{23} = \pi/2 - \theta_{23}^{CKM}$ is satisfied
with a good accuracy though it is not as precise as claimed in \cite{raidal}. 
We find $0.995 \leq \sin^2 2 \theta_{23} \leq 1.0$. 
The deviation from maximal mixing, 
\be
D_{23} = \cos \theta_{sun} |V_{cb}| \cos \delta = 0.035 \cos \delta, 
\label{d23_l1}
\ee
is relatively large at  $\delta \simeq 0$. 

\vspace{0.2cm}   

\noindent
4). The Jarlskog invariant equals 
\be
J_{lep} = - \frac{1}{2} \cos{\theta_{sun}} \sin^2{\theta_{sun}}
|V_{cb}| \sin{\delta} \sim -5 \times 10^{-3} \sin{\delta}.
\label{jlepton}
\ee
Its absolute value is larger than that in the neutrino 
scenario of sec.~3.1, but is an order of magnitude 
smaller than $J^{max}_{lep}$ (\ref{jmax}).

%%%%%%%%%%%%%%%%%%%%%%%%%%%%%%%%%%%%%%%%%%%%%%%%%%%%%%%%%%
\subsection{Hybrid scenario}
%%%%%%%%%%%%%%%%%%%%%%%%%%%%%%%%%%%%%%%%%%%%%%%%%%%%%%%%%%

The maximal 1-2 and 2-3 mixings may come from different 
mass matrices. To keep correct order of these rotations 
in the MNS matrix (\ref{gen1}), we have to 
assume that in the symmetry basis the maximal 1-2 mixing 
originates from the neutrino mass matrix,  
whereas the maximal 2-3 mixing is generated by the charged 
lepton mass matrix.

The CKM rotation can come from neutrinos 
or charged leptons and also mixed version is possible. 
We only discuss the former  two cases. 
In the first case, we have the CKM mixing from the neutrino mass matrix: 
\be
U_{\nu} = V^{CKM \dagger}  R_{12}^m, ~~~~~~
U_l = R_{23}^{m  \dagger}.
\label{rot-h}
\ee
For quarks we take equalities (\ref{quark-leps}) as in the 
``charged lepton'' scenario.  

This possibility looks more appealing than the second one. 
A  realization can be as follows. 
In the symmetry basis due to the quark-lepton  symmetry  
we have (\ref{up-nu}), $m_u = m_{\nu}^D$. 
This leads to the rotation which diagonalizes the neutrino Dirac 
mass matrix: 
\be
V_{\nu}^D = V_u = V^{CKM \dagger}. 
\ee
The maximal 1-2 rotation,  $R_{12}^m$, is the outcome of the seesaw
mechanism. It can be generated by  the  pseudo-Dirac (off-diagonal) 
1-2 structure of the Majorana mass matrix of the RH neutrinos~\cite{PS}. 
As a result, the rotation  matrix  (\ref{rot-h}) is reproduced. 
For the charged leptons and down quarks one should assume 
the lopsided scenario with a single maximal mixing.
%\be
%V_l = R_{23}^m, ~~~~~ V_d = I. 
%\ee\\
%
Here, the quark-lepton symmetry is broken.

In the second case, the CKM mixing comes from the charged leptons:
\be
U_{\nu} = R_{12}^m, ~~~~~~
U_l = V^{CKM}  R_{23}^{m  \dagger}.
\label{rot-hl}
\ee
%The CKM matrix is generated by the down quarks,  
%$V_d = V^{CKM}$, and one needs to assume that leptons have 
%an additional maximal 2-3 rotation. 
%
Both of the scenarios lead to the identical MNS matrix 
\be
U_{MNS}  = R_{23}^m V^{CKM \dagger} R_{12}^m  =
R_{23}^m \Gamma_{\delta}
R_{12}^{CKM \dagger}
R_{23}^{CKM \dagger}
R_{12}^m , 
\label{mns-h1}
\ee
where we have 
%introduced a CP phase $\delta$ in the same manner as before and 
ignored the $R_{13}^{CKM}$ rotation. 

Below we summarize the predictions 
of the hybrid scenario. 
The QLC relation (\ref{equality}) is satisfied to a good accuracy:
\be 
\sin \theta_{sun} &=& 
\sin \left( \frac{\pi}{4} - \theta_C \right) + 
\frac{1}{2 \sqrt{2}}
\sin \theta_C  |V_{cb}|^2, \\
%\ee
%(ignoring $O(|V_{cb}|^4)$ terms),  
%
%and consequently, 
%\be
\Delta \sin^2 \theta_{12} &=&  \frac{1}{\sqrt{2}}\sin \theta_{sun} \sin 
\theta_C |V_{cb}|^2  \simeq 1.4 \times 10^{-4}.  
\label{12-hb}
\ee
The 1-3 mixing angle is very small:  
\be
\sin \theta_{13} = \sin \theta_{C} |V_{cb}| \simeq 9.1 \times 10^{-3} 
\label{13-hb}
\ee
which corresponds  to $\sin^2 2\theta_{13} = 3.3 \times 10^{-4}$. 
The prediction for  $D_{23}$ reads 
\be
D_{23} = \cos \theta_{C} |V_{cb}| \cos \delta  \simeq 0.04 \cos \delta. 
\label{d23_hb}
\ee
It is  almost identical to  the one in the 
lepton bi-maximal scenario  (\ref{d23_l1}) but with 
replacing $\cos \theta_{sun}$ by $\cos \theta_{C}$. 
For the Jarlskog invariant we obtain 
\be
J_{lep} = \frac{1}{4}
\sin{\theta_{C}}  \cos{2\theta_{C}} |V_{cb}|
\sin{\delta} \simeq 2.1 \times 10^{-3} \sin{\delta}. 
\label{j_hb}
\ee

%%%%%%%%%%%%%%%%%%%%%%%%%%%%%%%%%%%%%%%%%%%%%%%%%%%%%%%
\section{Single maximal mixing}
%%%%%%%%%%%%%%%%%%%%%%%%%%%%%%%%%%%%%%%%%%%%%%%%%%%

To reproduce the QLC relation (\ref{equality}), 
it is sufficient to have a single maximal mixing 
in 1-2 rotation (sec.~2). 
%So, the bi-maximal 
%mixing is not the general feature  of models of lepton mixing. 
%
We discuss in this section the three scenarios 
which differ by the origin of large but not maximal atmospheric mixing. 
%We note that the non-maximal $\theta_{23}$ and the QLC relation opens a new 
%possibility of the Cabibbo angle as a unique fundamental parameter in fermion mixings.  

%%%%%%%%%%%%%%%%%%%%%%%%%%%%%%%%%%%%%%%%%%%%%%%%%%%%%%%
\subsection{Large 2-3 mixing from neutrinos}
%%%%%%%%%%%%%%%%%%%%%%%%%%%%%%%%%%%%%%%%%%%%%%%%%%%%%%%%%

Here we relax the assumption of maximal 2-3 mixing in the 
neutrino scenario considered in sec.~3.1.
The lepton mixing matrix is given by (\ref{mns-nu1}) with the replacement 
$R_{23}^m \rightarrow  R_{23}(\theta_{23}^{\nu})$, 
\be 
U_{MNS}  = V^{CKM \dagger}\Gamma_\delta  R_{23}(\theta_{23}^{\nu}) R_{12}^m .     
\label{mns-nu2}
\ee

Such a possibility can be realized in the following way.  
Suppose in the symmetry basis, (i) the up-quark mass matrix and the 
neutrino Dirac matrix are diagonal, (ii) the down quark matrix 
generates the CKM mixing: 
\be
m_u = m_{\nu}^D = diag, ~~~~ V_d = V_l = V^{CKM}, 
\ee
and (iii) the Majorana mass matrix of the right handed neutrinos 
has the following form  
\begin{equation}
M_R \approx \left(
\begin{tabular}{lll}
0 & $M_{12}$ & 0\\
$M_{12}$ & 0 & 0\\
0 & 0 & $M_{33}$\\
\end{tabular}
\right),
\label{eq:it}
\end{equation}
with $  M_{12}/M_{33} \geq  m_c^2/m_t^2$. 
Then, the see-saw mechanism leads to the maximal 1-2 mixing 
and enhancement of the 2-3 mixing~\cite{seesenh} when also  
non-zero but small 2-3 entries are introduced in (\ref{eq:it}). 
Typically the 1-3 mixing turns out to be very small, and  
an additional 1-3 rotation in 
the neutrino mixing matrix (\ref{mns-nu2}) can be neglected.

We first discuss constraints on $\theta^{\nu}_{23}$ 
from the CHOOZ and atmospheric neutrino data. 
Using 
$|U_{\mu 3}|^2 = \cos^2 \theta_{C}  
(\sin^2 \theta^{\nu}_{23} -
\sin 2\theta^{\nu}_{23} |V_{cb}| \cos \delta$) and 
the Super-Kamiokande allowed range \cite{hayato} 
gives a mild constraint 
$0.36 \leq \sin^2 \theta^{\nu}_{23} \leq 0.69$, or 
$37^{\circ} \leq \theta^{\nu}_{23} \leq 56^{\circ}$. 
The CHOOZ constraint is satisfied due to the relation (\ref{e3-mu3}).

Because of the non-maximal 2-3 mixing, 
the QLC relation is satisfied with slightly better accuracy 
as in the case of bi-maximal neutrino scenario of sec. 3.1. 
The correction to this relation reads 
\be 
\Delta \sin^2 \theta_{12} &=&  
%\sin^2 \left(\frac{\pi}{4} -\theta_C  \right) + 
\sin 2\theta_{C} 
\sin^2 \left(\frac{\theta^{\nu}_{23}}{2}\right) - 
\frac{1}{2} \sin^2 \theta_{C} 
\sin^2 \theta^{\nu}_{23} 
\nonumber \\
&-& 
\sin \theta_{C} \sin \theta^{\nu}_{23} 
\left(\cos \theta_{C} - \sin \theta_{C} \cos \theta^{\nu}_{23} \right)
|V_{cb}| \cos \delta.
\label{SC_bl_nu}
\ee
%
%where we have ignored the terms of the order $|V_{ub}|$ and $|V^2_{cb}|$.  
Neglecting the small $\delta$-dependent term 
in (\ref{SC_bl_nu}) 
%($\leq 4 \times 10^{-4}$),  
and using the bound on $\theta^{\nu}_{23}$, we obtain  
\be 
0.034 \leq 
\Delta \sin^2 \theta_{12}
\leq 0.079
\label{SC_deviat-nu2}
\ee
which corresponds to 
$2.2^{\circ}  \leq \theta_{sun} + \theta_C - \frac{\pi}{4} \leq 5.0^{\circ}$.

Since the scenario can accommodate the whole region of 
$|U_{\mu 3}|^2$ allowed by the present data, the deviation 
from maximal $\theta_{23}$,  
\be
D_{23} = 
\frac{1}{2} \cos 2\theta^{\nu}_{23} + 
\sin^2 \theta_C \sin^2 \theta^{\nu}_{23} - 
\cos^2 \theta_C |V_{cb}| \cos \delta ,  
\label{d23_bl_nu2}
\ee
can be large, $|D_{23}| \leq 0.16$, which gives the opportunity 
for verification in the next generation experiments. 
The Jarlskog invariant is enhanced by a factor of $\simeq 4.6$ 
in comparison with bi-maximal case, 
\be
J_{lep} =  \frac{1}{4}
\sin{2\theta_{C}}
\sin^3 \theta^{\nu}_{23}
|V_{cb}| \sin{\delta}
\leq
6.8 \times 10^{-3} \sin{\delta}
\ee
thanks to the mild constraint on $\theta^{\nu}_{23}$.\\

One can introduce small $\theta^{\nu}_{13}$ rotation into the 
bi-large matrix (\ref{mns-nu2}) of the order of the CHOOZ limit. 
%Ignoring $V_{cb}$ term We find that 
This gives an additional 
contribution to $\Delta \sin^2 \theta_{12}$ 
\be
\sin\theta^{\nu}_{13}  
\sin{\theta_{C}} \sin\theta^{\nu}_{23} 
\left(
\cos{\theta_{C}} - \sin{\theta_{C}} \cos\theta^{\nu}_{23} 
\right) \simeq 
0.1 \sin\theta^{\nu}_{13}  \sim \pm 0.016 
\ee
which can further reduce (for $\sin\theta^{\nu}_{13}  < 0$) the deviation from 
from the exact QLC relation.
Within the same approximation, $|U_{e3}|^2$ obtains an additional 
term of the order $\sin\theta^{\nu}_{13} $: 
\be
- \frac{1}{2}
\sin{2 \theta_{C}} \sin\theta^{\nu}_{23} 
\sin{2 \theta^{\nu}_{13}} \sim \mp 0.05
\ee
which mildly relaxes (tightens) the constraint on $\sin\theta^{\nu}_{23} $
for positive (negative) $\sin\theta^{\nu}_{13} $.

%%%%%%%%%%%%%%%%%%%%%%%%%%%%%%%%%%%%%%%%%%%%%%%%%
\subsection{Large 2-3 mixing from charged leptons }
%%%%%%%%%%%%%%%%%%%%%%%%%%%%%%%%%%%%%%%%%%%%%%%%%%%%%%%

One can relax the assumption of bi-maximal mixing also 
in the case of lepton scenario by introducing large 
but non-maximal $\theta^{l}_{23}$, so that the lepton mixing 
matrix takes the form 
\be
U_{MNS}  = R_{23}^l \Gamma_{\delta} R_{12}^m  V^{CKM \dagger}. 
\label{mns-l2}
\ee
The QLC relation is satisfied almost exactly and the correction  
(\ref{sun-l1}) 
%obtained in the lepton bi-maximal scenario 
remains unchanged.

Similar to the  $|U_{e3}|$-$|U_{\mu 3}|$ relation
in the neutrino-origin bi-large mixing scenario, there exists 
a relation 
\be
|U_{e3}|^2 = 
\tan^2 \theta^{CKM}_{23} |U_{e2}|^2 \simeq 
|V_{cb}|^2 \sin^2 \theta_{sun}
\label{e2-e3}
\ee
independent of $\theta^{l}_{23}$ and 
$\delta$. It immediately tells that $|U_{e3}|^2$ is small, 
$\simeq 5 \times 10^{-4}$.

Ignoring small $\delta$-dependent term, 
one can show that $\theta^{l}_{23}$ has a similar bound 
$36^{\circ} \leq \theta^{l}_{23} \leq 54^{\circ}$ 
as $\theta^{\nu}_{23}$ from atmospheric neutrino data (see sec.~4.1). 
So apparently the deviation from maximal 2-3 mixing  
\be
D_{23} = 
\frac{1}{2} \cos 2\theta^{l}_{23} + 
\cos \theta_{sun} \sin 2\theta^{l}_{23}
|V_{cb}| \cos \delta 
\label{d23_bl_l}
\ee
can cover whole region allowed by the Super-Kamiokande 
data, $|D_{23}| \leq 0.16$. 
%as in the neutrino-origin case. 
%
The Jarlskog invariant 
\be
J_{lep} = - \frac{1}{2}
\cos{\theta_{sun}} \sin^2{\theta_{sun}}
\sin 2\theta^l_{23}
|V_{cb}| \sin{\delta} 
\ee
being proportional to  
$\sin 2\theta^l_{23}$ is bounded by $J_{lep}$ (\ref{jlepton}) found for  
$\theta^{l}_{23} = \pi/4$.  
%CP violation is small independent of $\theta^{l}_{23}$ 
%in the lepton-origin bi-large mixing scenario.

One can introduce also small $\theta_{13}$ into the bi-large matrix 
(\ref{mns-l2}),  so that $|U_{e3}|$ saturates the CHOOZ limit. 
But, its effect to the QLC relation is $\sim$ 1\%, and 
it produces even smaller effect in $|U_{\mu 3}|$.

\vspace{0.2cm}

A non-maximal 2-3 mixing can also be introduced into the 
hybrid scenario described in sec. 3.3 by replacing 
$R_{23}^m$ by 
$R_{23}^l \equiv R_{23}(\theta^l_{23})$ 
in the MNS matrix in (\ref{mns-h1}). 
In this case, the correction to the QLC relation, (\ref{12-hb}), 
and the result for  $U_{e3}$ in Eq.  (\ref{13-hb}), are unchanged. 
The deviation parameter $D_{23}$ is 
given by that in the lepton-origin single maximal case  
(\ref{d23_bl_l}), but with replacement 
$\theta_{sun} \rightarrow \theta_{C}$. 
The upper bound on the deviation,  $|D_{23}| \leq 0.16$,  
remains unchanged.
The Jarlskog invariant gets an additional factor 
$\sin 2\theta^l_{23}$ in comparison with (\ref{j_hb}) .
%
%\be
%J_{lep} = \frac{1}{4}\sin{\theta_{C}}\cos{2\theta_{C}}\sin 2\theta^l_{23}
%|V_{cb}| \sin{\delta} \leq 2.1 \times 10^{-3}.
%\ee
%

%%%%%%%%%%%%%%%%%%%%%%%%%%%%%%%%%%%%%%%%%%%
\subsection{Large 2-3 mixing from neutrinos and charged leptons}
%%%%%%%%%%%%%%%%%%%%%%%%%%%%%%%%%%%%%%%%%%%%%%%%%%

The  large  2-3 mixing can appear as a sum of contributions from the
neutrinos and charged leptons. Let us assume that
as a result of the seesaw mechanism, the neutrinos
produce maximal 1-2 rotation and large but non-maximal 2-3
rotation in a way described in sec.~4.1. 
(Note that it is easier to get a single
maximal mixing from the seesaw mechanism.)   
The charged leptons generate the
CKM rotation and also relatively large (Cabibbo angle  size)
2-3 rotation.
So,
\be
U_{\nu} = R_{23}^{\nu}  R_{12}^m, ~~~~~~
U_l = V^{CKM \dagger} R_{23}^{l \dagger},
\label{rot-s}
\ee
and consequently,
\be
U_{MNS}  = R_{23}^{l} \Gamma_{\delta} V^{CKM \dagger}
R_{23}^{\nu} R_{12}^m.
\label{mns-s}
\ee
The difference from the neutrino scenario (sec. 4.1)
is that now the 2-3 rotation $R_{23}^{\nu}$ between 
$R_{12}^{CKM}$ and $R_{12}^m$ has the angle 
$\theta_{23}^{\nu}$ which is smaller than $\theta_{atm}$.
Therefore, the correction to the QLC relation (\ref{equality}) is smaller.
Instead of  (\ref{solangle}) we find, ignoring order  $|V_{ub}|$ terms, 
\be
\sin \theta_{sun} = \sin \left(\frac{\pi}{4} -\theta_C  \right) +
\frac{\sin \theta_C} {\sqrt{2}}
\left[1  - \cos (\theta_{23}^{\nu} - \theta_{23}^{CKM}) \right].
\label{solangle-s}
\ee
%where corrections of the order  $|V_{ub}|$ are neglected.

For the purpose of estimations of numbers we take, throughout 
this subsection, $\theta_{23}^{l} = \theta_{C} = 13^{\circ}$ and 
$\theta_{23}^{\nu} \simeq 2\theta_{C} = 27^{\circ}$.
The spirit behind the choice of these numbers is that we 
pursue the possibility that inherently there is no large mixing 
angle in building blocks of the MNS matrix.
The latter choice is also motivated as the smallest choice 
consistent with the large atmospheric angle.
Then, from (\ref{solangle-s}) we obtain $\theta_{sun} = 33^{\circ}$,
and $\sin^2 \theta_{sun} = 0.30$ which is substantially closer to the
central experimental value than the oscillation
parameter in the neutrino scenario.

The 1-3 mixing parameter determined now as 
\be
\sin \theta_{13} = \sin \theta_C 
\sin (\theta_{23}^{\nu} - \theta_{23}^{CKM}) 
\label{reangle-s}
\ee
has the mildly suppressed value in comparison with 
the neutrino-origin single maximal case (sec. 4.1):   
$\sin \theta_{13} = 0.093$, or $\sin^2 2\theta_{13} = 0.034$.

The 2-3 mixing matrix element is determined as
\be
U_{\mu 3} = \sin \left(\theta_{23}^{l} + \theta_{23}^{\nu} -
\theta_{23}^{CKM} \right) &+& 
2 \sin^2 \left(\frac{\theta_{C}}{2} \right)
%(1 - \cos \theta_C)
\cos \theta_{23}^{l} \sin (\theta_{23}^{\nu} - \theta_{23}^{CKM})
\nonumber \\
&+&
\sin \theta_{23}^{l} \cos (\theta_{23}^{\nu} - \theta_{23}^{CKM})
(e^{i \delta} - 1).
\label{atangle-s}
\ee
A notable feature of (\ref{atangle-s}) is that the argument 
of sine function (the first term in the RHS) is the addition of modest 
size angles, 
which make our ``no inherent large angle'' assumption 
in lepton mixing tenable.
In fact, under the assumption $\theta_{23}^{l} = \theta_{C}$, 
the 2-3 mixing angle can be written as 
\be
\sin^2 \theta_{23} &=& 
\sin^2 \left(\theta_{23}^{\nu} - \theta_{23}^{CKM} \right) +
\sin^2 \theta_{C} 
\left[1 - 
3 \sin^2 \left(\theta_{23}^{\nu} - \theta_{23}^{CKM} \right)
\right] 
\nonumber \\
&-&
\sin \theta_{C} \cos^2 \theta_{C} 
\sin 2\left(\theta_{23}^{\nu} - \theta_{23}^{CKM} \right)
\cos \delta, 
\ee
ignoring $\sin^4 \theta_{C}$ terms. 
Numerically, for $\theta_{23}^{\nu} = 27^{\circ}$, it gives  
$\sin^2 \theta_{23} \simeq 0.28 -0.16 \cos \delta$.
Therefore, the Super-Kamiokande bound is satisfied for 
$112^{\circ} \leq \delta \leq 248^{\circ}$.

The Jarlskog invariant can be written as 
\be
J_{lep} = \frac{1}{4}
\sin{\theta_{C}}
\sin 2\theta^l_{23}
\sin \left(\theta_{23}^{\nu} - \theta_{23}^{CKM} \right)
\left[
\cos{2\theta_{C}} + 
\sin^2{\theta_{C}}
\sin^2 \left(\theta_{23}^{\nu} - \theta_{23}^{CKM} \right)
\right]
\sin{\delta}. 
\ee
Numerically, keeping the same numbers as above, 
we obtain $J_{lep} = 9.1 \times 10^{-3}$, which is the largest among 
predictions from all the scenarios in this paper. It is because of the 
feature that some of the small angles in elements of 
the MNS matrix (\ref{mns-s}) are ``absorbed'' into the 
large angles, as in (\ref{reangle-s}) and (\ref{atangle-s}).

%%%%%%%%%%%%%%%%%%%%%%%%%%%%%%%%%%%%%%%%%%%%%%%%%%%%%%%%%%%%%%%
\section{Summary of the predictions by various scenarios}
%%%%%%%%%%%%%%%%%%%%%%%%%%%%%%%%%%%%%%%%%%%%%%%%%%%%%%%%%%%%%%%

\begin{table}
\vglue 0.5cm
\begin{tabular}{|c|c|c|c|c|}
\hline
       & \hspace{0.3cm} $\Delta \sin^2 \theta_{12}$ \hspace{0.3cm} & \hspace{0.3cm} $\sin^2 2\theta_{13}$ \hspace{0.3cm} &  $D_{23} \equiv \frac{1}{2}-s^2_{23}$ \hspace{0.0cm} & \hspace{0.3cm} $J_{lep}/\sin \delta$ \hspace{0.3cm} \\
Scenarios      &             &                      &        &      \\
\hline
neutrino bi-maximal (\ref{mns-nu1})& 0.051 & 0.10 $\pm$ 0.032 & 0.025  & $1.5\times 
10^{-3}$  \\
lepton bi-maximal (\ref{mns-l1})& $-6\times 10^{-4}$ & $2\times 10^{-3}$ & 0.035$^*$ & 
$5\times 10^{-3}$  \\
hybrid bi-maximal (\ref{mns-h1})& $1.4\times 10^{-4}$ & $3.3\times 10^{-4}$ & 0.04$^*$ & 
$2.1\times 10^{-3}$     \\
neutrino max+large (\ref{mns-nu2})& 0.057 $\pm$ 0.023  & 0.10 $\pm$ 0.032 & SK bound & $\leq 6.8\times 10^{-3}$  \\
lepton max+large (\ref{mns-l2})& $-6\times 10^{-4}$ & $2\times 10^{-3}$ & SK bound & $\leq 5\times 10^{-3}$   \\
hybrid max+large & $1.4\times 10^{-4}$ & $3.3\times 10^{-4}$ & SK bound & $\leq 2.1\times 10^{-3}$ \\
single maximal (\ref{mns-s})& 0.015  & 0.034 & $0.06 - 0.16$  &  $9.1\times 10^{-3}$  \\

\hline
\end{tabular}
\label{Tabzero}
\vglue 0.5cm

\caption[aaa]{
Predictions to the deviation from the QLC relation 
$\Delta \sin^2 \theta_{12}$, $\sin^2 2\theta_{13}$, the deviation 
parameter from the maximal 2-3 mixing $D_{23}$, and 
the leptonic Jarlskog factor $J_{lep}$ for different scenarios. 
The number in parenthesis in the first column indicates the equation 
number where the scenario is defined.
The uncertainties indicated with $\pm$ 
come from the experimental uncertainty of the  atmospheric 
mixing angle $\theta_{23}$. 
Whenever there exist uncertainty due to the CP violating phase 
$\delta$ we assume that $\cos \delta = 0$ to obtain an 
``average value''.  For the quantities 
 which vanish at $\cos \delta = 0$ (indicated by *) 
the numbers are calculated by assuming 
$\cos \delta = 1$ 
``SK bound''  implies 
the whole region allowed by the Super-Kamiokande:  
$|D_{23}| \leq 0.16$. 
The numbers for the last row (single-maximal case) are computed 
with the assumed values of $\theta^l_{23}=\theta_{C}$ and 
$\theta^{\nu}_{23}=27^{\circ}$. 
}

\vglue 0.2cm
\end{table}
%%%%%%%%%%%%%%%%%%%%% Table I %%%%%%%%%%%%%%%%%%%%%%%%%%%%%%%%%%%%%%%%%%%%%%%%%

We compare predictions of different scenarios and discuss perspectives 
to disentangle them. 
In the Table 1 we summarize predictions for observables obtained  
in  the last two sections. 
One can see some typical features 
of the predictions from various scenarios. 
The lepton and the hybrid scenarios 
can be characterized by extremely small deviation from 
the QLC relation, which may be unobservable experimentally. 
They  also have common features which predict small $\theta_{13}$ 
which probably requires facilities beyond the superbeam experiments.
These statements apply not only to bi-maximal scenarios 
but also to their variations with single maximal mixing angle.

On the other hand, the predictions of the ``neutrino" scenarios 
are markedly different. Both the bi-maximal and the single 
maximal cases predict relatively large 
deviation from the exact QLC relation of 
$\Delta \sin^2 \theta_{12}/\sin^2 \theta_{12} \sim 17$ \%. 
They lead to  relatively large $\theta_{13}$ just below the 
CHOOZ limit which will be detected by the next generation 
long-baseline and reactor experiments.

The neutrino (lepton and the hybrid) bi-maximal scenarios predict 
%(are consistent with) 
deviation from the maximal 2-3 mixing 
by 5-7 \%. The prediction is lost when we modify the scenario 
by allowing  the  (2-3) mixing to be  non-maximal.

There exists a relation characteristic to the neutrino scenario, 
$|U_{e3}| = \tan \theta_{C} |U_{\mu 3}|$, 
which holds independently of $\delta$ and of whether the 
neutrino-origin 2-3 angle is maximal or not.
Similarly, in the lepton scenario there exists an analogous 
relation $|U_{e3}| = \tan \theta_{C} |U_{e2}|$, which is 
again independent of whether the lepton-origin 2-3 angle 
is maximal or not.
They represent general consequences of the neutrino- and 
lepton-origin bi-large mixing scenarios, and can be tested 
by future measurement of $\theta_{13}$ as well as more precise 
determination of $\theta_{23}$ and $\theta_{12}$.

Throughout all scenarios, leptonic CP violation is small: 
the Jarlskog invariant is smaller than the presently allowed value 
by a factor of $\sim$ 10.

There exist simple relations between predictions of the lepton and the
hybrid scenarios.  For the deviation from the exact QLC equality we find 
\be \frac{(\Delta \sin^2 \theta_{12})_{l}} {(\Delta \sin^2
\theta_{12})_{h}} = \frac{\sqrt{2} \sin \theta_{sun}}{\sin \theta_C}
\simeq 3.4.  
\ee 
$\sin \theta_{13}$ and $D_{23}$ are related by 
\be
\frac{(\sin \theta_{13})_{l}}{(\sin \theta_{13})_{h}} = \frac{\sin
\theta_{sun}}{\sin \theta_C} \simeq 2.4, \nonumber \\
\frac{(D_{23})_{l}}{(D_{23})_{h}} = \frac{\cos \theta_{sun}}{\cos
\theta_C} \simeq 0.87.  
\ee 
However, it will be extremely difficult to measure the small values 
of $\theta_{13}$ and $D_{23}$, and consequently to check these relations. 
Therefore, distinguishing between these scenarios is an open question.

%quantities and therefore to distinguish between the two
%scenarios, except possibly for a precise determination of small
%$\theta_{13}$ in neutrino factories.

%%%%%%%%%%%%%%%%%%%%%%%%%%%%%%%%%%%%%%%%%%%%%%%%%%%%%%%%%%%%%%%
\section{Discussion and Conclusions}
%%%%%%%%%%%%%%%%%%%%%%%%%%%%%%%%%%%%%%%%%%%%%%%%%%%%%%%%%%%%%%%

To summarize, the current solar neutrino data shows a 
precise relation between the leptonic and the quark 1-2 mixing angles. 
The measured values of these angles sum up to $\pi/4$ in an 
accurate way such that the deviation of the central value is 
smaller than the experimental error at $1\sigma$ CL. 
The relation, which was referred as the QLC 
(quark-lepton complementarity) relation in this paper, 
seems indicative of a deeper connection between quarks and 
leptons, the most fundamental matter to date.

We have formulated general conditions under which the QLC 
relation is satisfied. They include:
(1) correct order of large rotations, which impose certain 
restrictions on  the neutrino and charge lepton mass matrices, 
(2) certain restrictions of CP-violating phases  
  in the mass matrices, and 
(3) absence of large renormalization group effects. 
We require that no other free parameter enters the relation 
between these angles, otherwise the relation implies the tuning 
of parameters.

We explored, first, a possibility that lepton mixings appear as 
the combination of maximal mixing  and the CKM rotations. 
This led to the ``bi-maximal minus CKM mixing'' scenario which 
has several different realizations.  
These realizations differ by the ways of how maximal mixings are generated. 
The generic prediction of all these realizations is very small deviation 
of 2-3 mixing from maximal. So that if large deviation 
is observed the scenario will be excluded.

Natural possibility would be the neutrino origin of the bi-maximal 
structure.  It leads to the QLC relation only at an approximate level, 
which is consistent with the current experimental data. 
This scenario can be identified by relatively large 1-3 mixing 
which is close to the present upper bound. 
In the (charged) lepton-origin and hybrid bi-maximal scenarios, deviation 
from the QLC relation, the 1-3 mixing angle, and deviation of 
the 2-3 mixing angle from the maximal one are predicted to be all 
very small. The former two features are shared by their 
bi-large extension, but the last one not.

Let us  make several theoretical and heuristic remarks:

\vspace{0.1cm}
\noindent
1). We have considered the origin of lepton mixing as 
the ``maximal mixing minus Cabibbo mixing''. 
There are two problems in this context: 

\begin{itemize}

\item

the origin of  maximal (or bi-maximal mixing), 

\item

propagation of the Cabibbo (or CKM-) mixing 
to the leptonic sector.

\end{itemize}

The latter is rather non-trivial especially for the first and 
the second generation fermions in view of a large difference in 
mass hierarchies: 
$m_{e}/m_{\mu} = 0.0047$ and $m_{d}/m_s = 0.04 - 0.06$ as well as  
difference in masses of the  s-quark and muon. 
The precise quark-lepton symmetry should show up in mixing and not 
in mass eigenvalues. This can be done rather easily in the two 
generation context but 
difficult to implement for the first and second families 
in the three generation case~\cite{mina-smi}.

So, the main problem is propagation of the Cabibbo (or CKM)
mixing from the quark sector to the lepton sector. Since  
the quark-lepton symmetry is broken by masses of quarks and lepton, 
one does not expect that the quark mixing is ``transmitted'' to
the lepton sector exactly. On general ground one would 
get corrections to the mixing angle of the order
\be
\Delta \theta_{12} \sim  \theta_C \frac{m_d}{m_s} \sim 
0.5^{\circ} - 1^{\circ}
\label{corr-m}
\ee
which,  however, is below the present $1\sigma$ accuracy.

For illustration let us outline one possible scenario of
such a propagation of mixing in the case of neutrino origin of
maximal 1-2 mixing.

(i). The first and the second generation of fermions form the doublet of
the flavor group and acquire masses independently of the third
generation (singlet of the group). This is required
to reconcile the propagation of the Cabibbo
mixing with the  $b - \tau$ unification.

(ii). The quark-lepton symmetry leads to the approximate equality of
matrices of
the Yukawa couplings for the first and the second generations.
To explain the difference of masses of muon and $s$-quark  at GUT
scale one needs to introduce
two different Higgs doublets with different VEV's for quarks and for leptons.
Notice that $m_s \approx m_{\mu}$ at the electroweak (EW) scale, so 
that if
the flavor symmetry is realized at the EW scale one Higgs doublet is
sufficient.
In this case however the problem of flavor changing neutral currents both
in the lepton and quark sectors becomes very severe.

(iii). In the basis where the Dirac mass matrices of up-quarks and
 neutrinos are diagonal the 
matrices of the Yukawa couplings of the down quarks and charged leptons
should be nearly equal and  singular to reconcile
equal mixings and different mass hierarchies of the
quarks and leptons. The singularity and quark-lepton symmetry are
broken by terms of the order $m_d/m_s$ and  this
leads  to the correction given in (\ref{corr-m}).

We emphasize that what is really needed for the QLC relation 
to hold is the single maximal mixing in the 1-2 rotation either 
from neutrino or from lepton sectors. 
Theoretically, the single maximal mixing can be 
realized much more easily. 
The mass matrix of the RH neutrinos can be the origin of the 
maximal mixing for the first and the second generations and 
it can lead to enhancement of the 2-3 mixing.

\vspace{0.1cm}
\noindent
2). It is not excluded that the quark-lepton connection,  
which leads to relation between the angles,  is not so direct.  
It may work for the Cabibbo angle only, since 
$\sin\theta_C$ may turn out to be a generic parameter 
of the whole theory of the fermion masses.   
Therefore, it may appear in various places as the mass ratios and the 
mixing angles. An empirical relation 
\be
\sin \theta_C \approx \sqrt{\frac{m_\mu}{m_\tau}}
\ee
is in favor of this point of view.

\vspace{0.1cm}
\noindent
3).  One can consider some variations of the QLC equality 
(\ref{equality}). Noting that the 2-3 leptonic mixing angle 
measured with the atmospheric neutrinos is nearly maximal,   
$\theta_{atm} \equiv \theta_{23} \simeq \pi/4$,  
we may write instead of (\ref{equality})  
\begin{equation}
\theta_{sun} + \theta_C = \theta_{atm}, 
\label{sumatm}
\end{equation}
allowing possible extension to the case of non-maximal 
$\theta_{atm}$.

\vspace{0.1cm}
\noindent
4). Still the QLC relation can be accidental.  
There is also another non-trivial coincidence:  
\begin{equation}
\theta_{sun} + \theta_{\mu \tau} = \frac{\pi}{4}~,          
\label{sum2} 
\end{equation}
where the angle $\theta_{\mu \tau}$ is determined by the equality  
\begin{equation}
\tan \theta_{\mu \tau} \approx \sqrt{\frac{m_\mu}{m_\tau}}.
\label{angmutau}
\end{equation}

Apparently, the equalities (\ref{sumatm}) and (\ref{sum2}) have different 
interpretations from 
the QLC relation. 
In particular, (\ref{sum2}) is a pure leptonic relation.

\vspace{0.1cm}
\noindent
5).  The most important future measurements turn 
out to be:

\noindent
(i) Precise measurements of the 1-2 leptonic mixing and further checks of 
the QLC relation. The accuracy in $\sin^2 \theta_{sun}$ determination 
must be better than 10\% to discriminate the neutrino version 
of scenario. 

\noindent
(ii) Searches for deviation of the 2-3 mixing from the maximal one 
which can discriminate whole ``bi-maximal minus CKM'' approach.   

\noindent
(iii) Measurements of the 1-3 mixing angle.  

\vspace{0.1cm}

In conclusion, it is possible that the equality (\ref{equality})  
is  not accidental, thus 
testifying for a certain quark-lepton relation. 
%
%we can not exclude that the equality (\ref{equality}) 
%is not accidental thus testifying for certain  quark-lepton relation. 
Implementation of the equality naturally involves the idea that 
the lepton mixing appears as maximal mixing minus the Cabibbo mixing. 
In this sense, the quark and lepton mixings are complementary. 
The approach leads to a number of interesting relations between 
the lepton and quark mixing parameters which can be tested in future 
precision measurements.

\section{Acknowledgments}

One of us (A. Yu. S.) is  grateful to  M. Frigerio   
for fruitful discussions. 
This work was supported by the FY2004 JSPS Invitation 
Fellowship Program for Research in Japan, S04046, 
and by the Grant-in-Aid for Scientific Research, No. 16340078, 
Japan Society for the Promotion of Science.

%\section{First Appendix}

\renewcommand{\theequation}{A.\arabic{equation}}

%%%%%%%%%%%%%%%%%%%%%%%%%%%%%%%%%%%%%%%%%%%%%%%%%%%%%%%%%%%%%%%%%%%%%%%%%%%%%%%%%%%%%%%%%%%%%%%%

%\balance

\end{document}